%% file: IEEE-conference-template-062824/main.tex
\documentclass[conference]{IEEEtran}

\usepackage{cite}
\usepackage{amsmath,amssymb,amsfonts}
\usepackage{algorithmic}
\usepackage{graphicx}
\usepackage{textcomp}
\usepackage[table]{xcolor}
\usepackage{xcolor}
\usepackage{orcidlink}
\usepackage{hhline, multirow}

\def\BibTeX{{\rm B\kern-.05em{\sc i\kern-.025em b}\kern-.08em
    T\kern-.1667em\lower.7ex\hbox{E}\kern-.125emX}}
\begin{document}

\newcommand{\api}[1]{\textcolor{black}{\textit{TOSense}}}

\title{Demo: \textit{TOSense} – What Did You Just Agree to?\\
}

\author{
Xinzhang Chen\,\orcidlink{0009-0005-8901-2618},
Hassan Ali\,\orcidlink{0000-0002-1701-0390},
Arash Shaghaghi\,\orcidlink{0000-0001-6630-9519},
Salil S. Kanhere\,\orcidlink{0000-0002-1835-3475},
Sanjay Jha\,\orcidlink{0000-0002-1844-1520} \\
School of Computer Science and Engineering, The University of New South Wales, Sydney, Australia \\ \texttt{\{xinzhang.chen,hassan.ali,a.shaghaghi,salil.kanhere,sanjay.jha\}@unsw.edu.au}}

\maketitle

\vspace{0.5em}
\begin{center}
\textit{This is the preprint version of a paper accepted at IEEE LCN 2025. Please cite the final published version.}
\end{center}
\vspace{0.5em}

\begin{abstract}
Online services often require users to agree to lengthy and obscure Terms of Service (ToS), leading to information asymmetry and legal risks. This paper proposes \api{}---a Chrome extension that allows users to ask questions about ToS in natural language and get concise answers in real time. The system combines (i) a crawler ``tos-crawl'' that automatically extracts ToS content, and (ii) a lightweight large language model pipeline: MiniLM for semantic retrieval and BART-encoder for answer relevance verification. To avoid expensive manual annotation, we present a novel Question Answering Evaluation Pipeline (QEP) that generates synthetic questions and verifies the correctness of answers using clustered topic matching. Experiments on five major platforms, Apple, Google, X (formerly Twitter), Microsoft, and Netflix, show the effectiveness of \api{} (with up to 44.5\% accuracy) across varying number of topic clusters. During the demonstration, we will showcase \api{} in action. Attendees will be able to experience seamless extraction, interactive question answering, and instant indexing of new sites.
\end{abstract}


\section{Introduction}
\label{s:introduction}
Online services generally require users to accept Terms of Service (ToS) (including End User License Agreement (EULA) and Privacy Policy) before use. These ToS describe the rights and obligations between the platform and users, but are often lengthy and frequently use legal jargons~\cite{zaeem2018privacycheck, mcdonald2008cost}, making them inaccessible to most users. McDonald and Cranor~\cite{mcdonald2008cost} estimated that if American netizens were to read all such encountered policies, it would cost 201 hours annually ($\sim$\$781 billion worth of lost productivity). Empirical studies show that the vast majority of users accept ToS without meaningful review, 86\% spent less than one minute reading, and 93\% accepted it despite the inclusion of intentionally unreasonable clauses~\cite{Obar02012020}. This ``default consent'' phenomenon causes users to unknowingly accept legally binding terms such as ultimate ownership of the account, automatic renewal, and unilateral removal of posted content.

To address this problem, a variety of systems have attempted to improve the accessibility and transparency of ToS~\cite{tosdr, lippi2019claudette}. Most of these systems rely on recent developments in Large Language Models (LLM) to let users interactively understand ToS~\cite{lippi2019claudette, zaeem2018privacycheck, harkous2018polisis}. However, to the best of our knowledge, almost all of the current systems suffer from several limitations:
\textit{(A) Limited Coverage:} Many systems are limited to either predefined platforms~\cite{tosdr} or predefined topics (for example, mainly limited to privacy/security~\cite{zaeem2018privacycheck, harkous2018polisis}). \textit{(B) Ineffectiveness:} These systems often make impractical assumptions regarding document structure and the level of user engagement, rendering them ineffective for practical scenarios. For example, Claudette requires its users to manually paste ToS content on a separate portal everytime they encounter a ToS~\cite{lippi2019claudette}. Another key limitation is to assume that ToS have a single-page structure that makes it difficult to handle complex organization of clauses linked across multiple pages. \textit{(C) Lack of Comprehensive Evaluations:} Due to the absence of suitable benchmarks, it is significantly challenging to evaluate the reliability of such systems~\cite{harkous2018polisis}.

To address these challenges, we present \api{}---a Chrome extension combined with the LLM pipeline to help users interactively read and understand ToS---along with a Question/Answer Evaluation Pipeline (QEP) to quantify the reliability of \api{} without relying on labor-intensive manual annotations of ToS.
Our novel contributions are:
\begin{itemize}
    \item We implement \texttt{tos-crawl}, a crawler that supports automatic identification and extraction of multi-page ToS content. It supports features such as language filtering, login redirection avoidance and version deduplication.

    \item We present a novel Question/Answer Evaluation Pipeline (QEP) specifically suited for ToS that employs T5 question generator model along with a clustering algorithm to quantify the question answering capability of \api{} without relying on manually curated datasets.
\end{itemize}

\section{System Overview and Architecture}

\label{s:system-overview}
\begin{figure}
    \centering
    \includegraphics[width=0.85\linewidth]{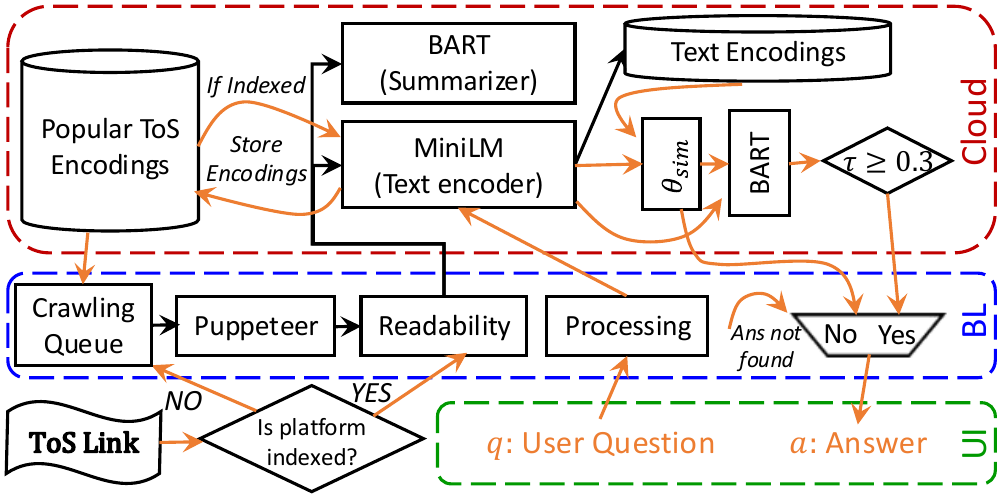}
    \caption{System architecture of \api{}}
    \label{fig:illustration}
\end{figure}

\api{} comprises three layers: the Cloud Layer, the Backend Layer, and the User Interface. These layers are illustrated in Fig.~\ref{fig:illustration} and are detailed below.

\noindent \textbf{A) The Cloud Layer:}
The cloud layer of \api{} is its main computational unit. Cloud layer hosts two LLMs as a service: MiniLM and BART~\cite{lewis2019bart}. \api{} uses MiniLM to encode textual input into feature vectors as shown in Fig.~\ref{fig:illustration}. When a user query is received, \api{} asks MiniLM to first encode the user query and then uses cosine similarity (denoted by $\theta_{sim}$ in Fig.~\ref{fig:illustration}) to retrieve an answer from the ToS that shows the highest cosine similarity with the encoded query. The aforementioned pipeline would output an answer irrespective of whether or not the user query is relevant to the ToS. To address such a case, \api{} performs a hypothesis test on the query and the identified answer using BART. BART outputs a number $b \in [0, 1]$ quantifying the relevance between the query and the answer. If $b < \tau$ \api{} responds with a predefined statement that it was unable to find an appropriate answer for the user query, and simply prints the answer if $b \geq \tau$. We heuristically set $\tau = 0.3$ as it worked best for our experiments.

\textit{Maintaining Popular ToS Base:} To ensure freshness and efficiency, the cloud layer periodically appends the ToS links of previously indexed platforms to the backend crawling queue, automatically triggering the re-crawl process. The updated documents are passed through the MiniLM encoder, and stored back to the Popular ToS Encodings (see Fig.~\ref{fig:illustration}). When users visit these indexed platforms, the system can directly retrieve the pre-processed encodings results to perform semantic search and question-answering without real-time crawling, greatly reducing response latency.

\noindent \textbf{B) The Backend Layer:}
The backend layer performs server-side crawling and extraction of ToS. It maintains a crawling task queue, which contains the platform records to be processed. A separate worker process continuously polls the queue and automatically calls the crawler module to process the platforms in the task. The crawler module is implemented based on the Puppeteer framework, augmented with the Stealth plug-in to circumvent anti-crawling mechanisms, and supports simulating user behaviors, such as automatically scrolling pages, expanding hidden content, and skipping login redirects. The module has recursive link tracking capabilities, can identify and extract clause-related subpages, and filter out the latest versions based on built-in rules. After the page is loaded, the content will be cleaned through the Mozilla Readability algorithm and converted to Markdown format, and finally sent to the cloud layer together with metadata for semantic encoding and index updates.

\noindent \textbf{C) The User Interface (UI) Layer:}
The UI layer is the main interactive entry point for users and is implemented as a Chrome browser extension based on the Plasmo framework. When a user visits any website, the extension will first detect whether the platform has been indexed by the system. If it has been indexed, the extension notifies the user via a sidebar popup and support direct Q\&A interaction based on the preprocessed ToS content. If the current platform has not been indexed, the extension will automatically analyze the page content to detect potential links related to ToS. When such links are detected, the extension prompts the user with the option to add the platform to the crawling queue. Upon confirmation, the link is appended to the crawling queue and will be indexed in the future. Once indexed, future visits to the same platform will enable direct ToS interaction without requiring repeated crawling.

\begin{figure}
    \centering
    \includegraphics[width=0.85\linewidth]{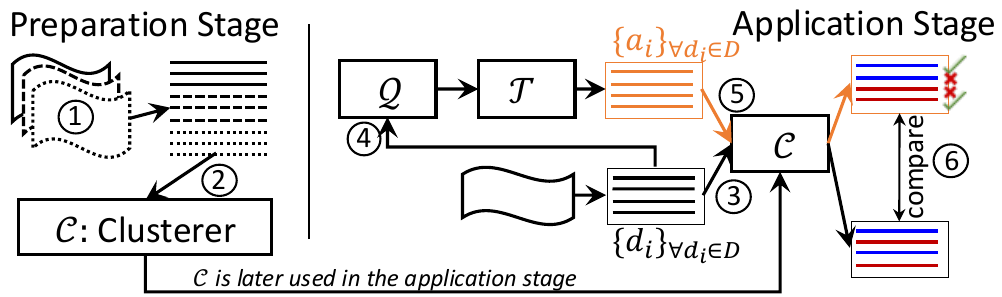}
    \caption{Novel Question/Answer Evaluation Pipeline (QEP) with $k$=2 clusters.}
    \label{fig:qep_evaluation}
\end{figure}
\section{Evaluation}
\label{s:evaluation}

As discussed previously, the lack of benchmarks presents a major challenge towards comprehensively evaluating the performance of Question/Answer (QA) models for ToS. In this section, we first present a novel QA Evaluation Pipeline (QEP) that addresses this challenge and then report results of our evaluation on 5 popular ToS platforms.

\noindent \textbf{Evaluation Methodology:}
Formally, we denote \api{} as $\mathcal{T}$ that takes as inputs a ToS $D$ along with a question $q$ and returns the answer $a=\mathcal{T}(q|D) \in D$. Our objective is to tell whether or not $a$ correctly answers $q$ without relying on a curated dataset. QEP achieves this in the following steps:

\noindent \underline{(Step 1):} QEP maintains a set of ToS (see step 1 in Fig.~\ref{fig:qep_evaluation}) from popular platforms such as Facebook and Google. In the preparation phase, QEP first encodes statements of these ToS into feature vectors by processing them through the MiniLM QA model. These feature vectors represent the semantics of each statement of ToS. QEP then uses the $k$-means clustering algorithm $\mathcal{C}$ to group these feature vectors into $k$ clusters (step 2 in Fig.~\ref{fig:qep_evaluation}). The intuition behind the clustering is that each of these $k$ learned clusters would represent a unique topic.
    
\noindent \underline{(Step 2):} In the application phase, given a ToS $D$, QEP first assigns each statement $d_i \in D$ to one of the $k$ previously identified clusters (see step 3 in Fig.~\ref{fig:qep_evaluation}). QEP then employs a T5-based automated question generator\footnote{Huggingface token: iarfmoose/t5-base-question-generator} $\mathcal{Q}$ to generate a question for each statement (step 4). Formally, $q_i=\mathcal{Q}(d_i),\ \forall d_i \in D$.
    
\noindent \underline{(Step 3):} Given $D$ and $q_i$, QEP invokes \api{} $\mathcal{T}$ to identify the answer $a_i = \mathcal{T}(q_i|D)$ from $D$. To check whether or not $a_i$ correctly answers $q_i$, QEP compares the clustering labels of $a_i$ and $d_i$ (steps 5 and 6 in Fig.~\ref{fig:qep_evaluation}). If $\mathcal{C}(a_i) = \mathcal{C}(d_i)$---i.e., the answer identified by \api{} is assigned the same cluster as the statement used to generate $q_i$ in step 4 of Fig.~\ref{fig:qep_evaluation}---QEP marks the answer to be correct.

\input{IEEE-conference-template-062824/tables/table_acc}

\noindent \textbf{Results:}
We maintain a set of ToS from 15 different platforms---Adobe, Apple, Coursera, Dropbox, Epic Games, Facebook, GitHub, Google, Instagram, OpenAI, Quora, Reddit, Stream, TikTok and X. ToS statements of these platforms are used in the QEP preparation phase (see Fig.~\ref{fig:qep_evaluation}) to train the clustering algorithm. 
For this experiment, we use $k \in \{5, 10, 15, 20, 30, 50, 80\}$ and report results in Table~\ref{tab:entangled_summaries_results_global}. We observe that as $k$ increases, the accuracy of \api{} drops slightly. This is expected as increasing $k$ also increases the number of output clusters to which $a_i$ might be assigned, leading to a slight decrease in accuracy. However surprisingly this decrease is not as sharp as one might expect. For example, the drop in accuracy is 2.5\% for Apple and 7\% for Google in Table~\ref{tab:entangled_summaries_results_global}. This shows that \api{} gives a fairly stable performance under multiple evaluation setups of QEP.

\input{IEEE-conference-template-062824/tables/topic_comparison}
We also compare the distribution of topics in ToS of multiple platforms. To achieve this, we manually assign a topic to each of the identified clusters in step 2 of Fig.~\ref{fig:qep_evaluation} by analyzing the statements grouped into each cluster. We then report the total number of statements from each ToS assigned to each topic (or cluster) in Table~\ref{tab:topic_comparison}. We observe a similarity in the proportion of topics discussed by all 5 platforms with a few exceptions. For example, we observe that X focuses notably more frequently on legally binding end-users to the ToS and relatively less frequently on privacy and security aspects as compared to other platforms. In contrary, Netflix focuses most frequently on the proprietary ownership of its contents as compared to other platforms.

\section{System Deployment and Performance Evaluation}
\label{s:system-deployment-and-performance-evaluation}
To evaluate the availability and operational efficiency of the \api{} system under actual deployment conditions, we deployed the complete system on a lightweight virtual machine running Ubuntu 24.04, which was allocated 2 virtual CPU cores (Intel Xeon E5-2683 v4) and 8GB of RAM, with no GPU acceleration. We then selected the same five representative platforms as Table~\ref{tab:entangled_summaries_results_global}, assuming these platforms were indexed, and sentence-level embeddings were pre-generated and cached in the system's ToS database. This setup simulates a typical stable state in a real-world deployment, where the query phase involves only semantic retrieval and answer generation, without any regeneration of embeddings.

For each platform, we executed five equivalent query requests using the question ``Does this service share my data with third parties?'' and recorded the following three metrics:

\begin{itemize}
    \item \textbf{Latency:} The total time from client request to response, used to assess user-perceived responsiveness.
    
    \item \textbf{Resource Usage (CPU and Memory):} System-level CPU and memory usage, recorded using the \texttt{psutil} tool, immediately after each query.
    
    \item \textbf{Timing:} The actual execution time within the server, encompassing only the semantic retrieval and answer verification phases (MiniLM + BART).
\end{itemize}

\input{IEEE-conference-template-062824/tables/performance_evaluation}

The results show that even without using a GPU, the \api{} system can still keep query response latency below 3 seconds on all five platforms. CPU utilization for each query remained stable between 91\% and 93\% and memory usage remained below 38\% across all platforms, demonstrating good resource stability.

\section{What Will Be Demonstrated}
\label{s:what-will-be-demonstracted}
During the demonstration, we will showcase the complete workflow of \api{} in different usage scenarios with the attendees, and show how it can help users understand complex ToS content and identify potential risks. We  acknowledge that no attendees' identity information will be collected during the demonstration, and all interactions will be completed anonymously. In contrast to conventional LLM-based tools, such as ChatGPT, which rely on manual copy-paste of relevant documents, \api{} provides an automatic, end-to-end ToS understanding solution that covers the entire pipeline from document extraction to semantic question answering. Attendees will be able to experience the following core functionalities:

\begin{itemize}
    \item \textbf{Install the Plugin and Trigger Platform Detection:} During the demonstration, attendees will be able to download and load the \api{} Chrome extension into their local Chrome browser. As they browse any website, the extension will automatically determine whether the platform has been indexed: If it has, it enters interactive Q\&A mode; otherwise, it will first analyze whether the current page contains potential ToS related links. If a suitable link is detected, it prompts the user to queue the platform for crawling.

    \item \textbf{Query the ToS and Receive Semantic Answers:}  For the indexed platform, attendees can issue natural language questions, such as: ``Can I permanently delete my account and data?'', ``What personal data does this platform collect?'' through the extension. The system will retrieve relevant content from the platform's ToS and display the answer after relevance verification; if no suitable answer is found, the system will explicitly notify the user that no valid answer could be found within the document to avoid misleading.

    \item \textbf{Observe Automated Crawling and Document Processing:} When a previously unindexed platform is submitted for processing, the \texttt{tos-crawl} module is triggered to recursively follow identified ToS-related links. Attendees will observe how it simulates user interactions such as auto-scrolling, expanding collapsible sections, and bypassing login prompts to extract content. For content that spread across multiple pages, the module automatically cleans the content, deduplicates versions, and consolidates the structure, ultimately generating a unified Markdown document for downstream semantic encoding. This processed result is cached, so that future visits to the same platform can skip crawling and enable immediate interaction.
\end{itemize}

\begin{figure}[h]
    \centering
    \includegraphics[width=0.8\linewidth]{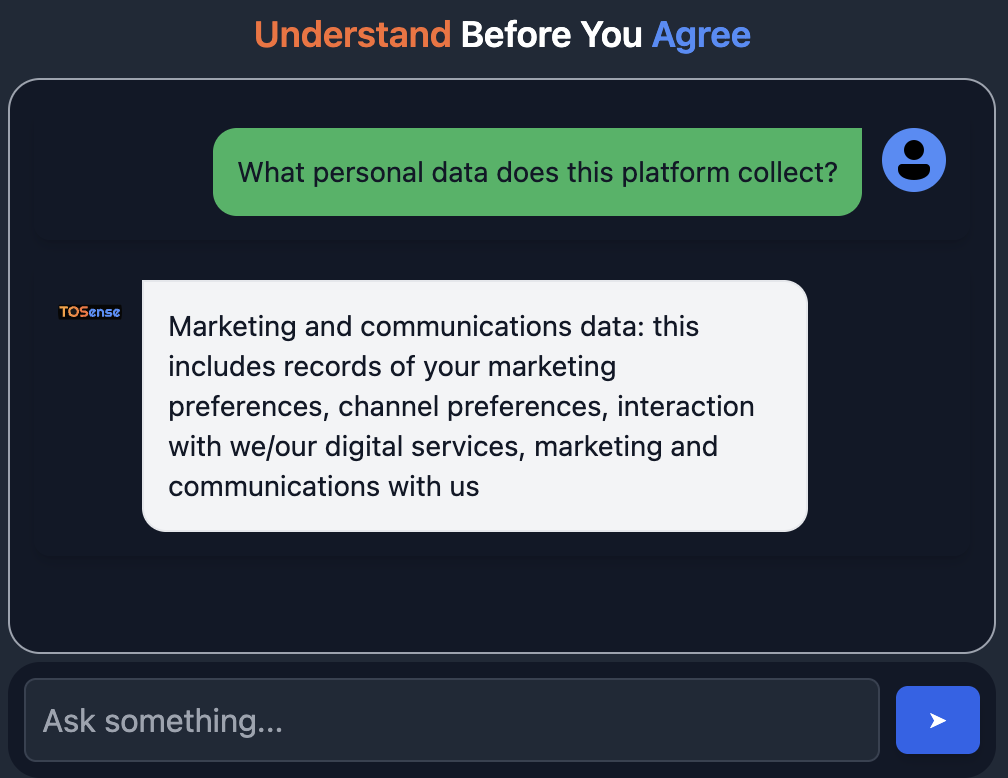}
    \caption{TOSense plugin interface responding to a personal data query on an indexed platform.}
    \label{fig:plugin-screenshot}
\end{figure}

\vspace{1em}
\noindent Fig.~\ref{fig:plugin-screenshot} illustrates the \api{} plugin answering a user’s query about personal data collection, based on the extracted ToS content.

\section{Limitations and Future Work}
Although \api{} provides a usable prototype and demonstrates the feasibility of intelligent interpretation of ToS based on LLM, this is only the first step towards comprehensive understanding. The following limitations remain, suggesting concrete directions for future improvement.

\noindent \textbf{(1) Crawler Robustness:} The current crawler uses basic interactions such as automatic scrolling and click-to-expand, as well as fixed keywords (such as ``terms'', ``privacy'', ``policy'', etc.) and heuristic rules to filter links. For ToS pages that use single-page application routing, asynchronous API loading, or use unconventional URL naming, this approach is prone to missing or only crawling part of the content. In the future, we plan to integrate lightweight computer vision techniques to detect related components and adopt semantic matching using pre-trained language models to identify likely ToS links.

\noindent \textbf{(2) Lack of User-Centred Evaluation:} Currently, no systematic user research has been conducted to evaluate whether \api{} can improve users' understanding of terms or change ``default consent'' behavior in the long term. The actual utility still needs further empirical exploration, we plan to conduct  A/B testing and task-based comprehension studies, focusing on factors such as answer usefulness and behavioral changes (such as click-through rate, summary reading time, etc.) to verify user benefits in real browsing scenarios.

\noindent \textbf{(3) Answer Accuracy and Model Limitations:} The current system has achieved a maximum accuracy of 44.5\% on the Microsoft ToS dataset (Table~\ref{tab:entangled_summaries_results_global}). While this demonstrates the feasibility of the system's question-answering capabilities, it also highlights the limitations of using lightweight language models (such as MiniLM and BART) for semantic retrieval and answer relevance in the context of ToS. For the future work, we plan to explore instruction-tuned models that better follow natural language queries, and apply domain-adaptive pretraining using large-scale corpora of ToS documents to enhance the model’s understanding of ToS-specific structure and terminology.

\section{Conclusion}
\label{s:conculsion}
In this paper we present \api{}---a Chrome extension that combines a ToS crawler with the MiniLM + BART-encoder LLM pipeline---that answers user queries about ToS contents in real time. We also propose a novel pipeline to quantitatively evaluate \api{} without manual annotation. Experiments on five selected platforms show that \api{} maintains a stable accuracy rate at different clustering hyperparameters. During the demonstration, attendees will be able to observe \api{}’s core interaction flow and experience the system hands-on. The adaptability of the crawler and user research still need to be improved, but \api{} has laid a reproducible technical foundation for improving the transparency of legal texts and promoting user informed consent. A demonstration build of \api{} is available at https://xinzhang-chen.github.io/TOSense-Landing-Page/, which includes the \api{} Chrome extension, \texttt{tos-crawl} module source code, usage instructions and demonstration videos.

\bibliographystyle{IEEEtran}
\bibliography{IEEE-conference-template-062824/bibliography/bib}

\end{document}

%% file: IEEE-conference-template-062824/tables/table_acc.tex
\begin{table}[t]
\scriptsize
\centering
\caption{
    Effect of cluster size ($k$) on QEP accuracy across five selected platforms.
}
\resizebox{0.9\linewidth}{!}{%
\begin{tabular}{l|c|c|c|c|c|c|c}
\hline
\multirow{2}{*}{\textbf{Platform}} & \multicolumn{7}{c}{\textbf{Number of Clusters:} $k$} \\
\cline{2-8}
& \textbf{5} & \textbf{10} & \textbf{15} & \textbf{20} & \textbf{30} & \textbf{50} & \textbf{80} \\
\hline

Apple & 0.275 & 0.245 & 0.245 & 0.270 & 0.260 & 0.220 & 0.250 \\

Google & 0.335 & 0.310 & 0.280 & 0.275 & 0.275 & 0.265 & 0.265 \\

X & 0.260 & 0.200 & 0.185 & 0.220 & 0.190 & 0.185 & 0.185 \\

Microsoft & 0.445 & 0.395 & 0.325 & 0.385 & 0.370 & 0.325 & 0.320 \\

Netflix & 0.330 & 0.285 & 0.245 & 0.240 & 0.250 & 0.250 & 0.225 \\
\hline

\end{tabular}%
}
\label{tab:entangled_summaries_results_global}
\end{table}

%% file: IEEE-conference-template-062824/tables/topic_comparison.tex
\begin{table}[t]
\scriptsize
\centering
\caption{
   Topic distribution in ToS documents across five selected platforms
}
\resizebox{1\linewidth}{!}{%
\begin{tabular}{l|c|c|c|c|c}

\hline
\multirow{3}{*}{\textbf{Platform}} & \multicolumn{5}{c}{\textbf{Topics}} \\
\cline{2-6}
& \textbf{Third-Party} & \textbf{Proprietary} & \multirow{2}{*}{\textbf{Privacy}} & \textbf{Legally} & \multirow{2}{*}{\textbf{Security}} \\
& \textbf{Services} & \textbf{Rights} & & \textbf{Binding} & \\
\hline

Apple & 0.13 & 0.26 & 0.18 & 0.24 & 0.19 \\

Google & 0.15 & 0.30 & 0.18 & 0.16 & 0.21 \\

X & 0.11 & 0.24 & 0.09 & 0.42 & 0.14 \\

Microsoft & 0.28 & 0.30 & 0.11 & 0.16 & 0.15 \\

Netflix  & 0.13 & 0.33 & 0.11 & 0.19 & 0.24 \\
\hline

\end{tabular}%
}
\label{tab:topic_comparison}
\end{table}

%% file: IEEE-conference-template-062824/tables/performance_evaluation.tex
\begin{table}[t]
\scriptsize
\centering
\caption{Runtime performance of TOSense on five selected platforms.}
\resizebox{1\linewidth}{!}{%
\begin{tabular}{l|c|c|c|c}
\hline
\textbf{Platform} & \textbf{Latency (s)} & \textbf{CPU (\%)} & \textbf{RAM (\%)} & \textbf{Timing (s)} \\
\hline
Apple      & 2.15 & 91.84 & 35.42 & 2.080 \\
Google     & 2.14 & 92.24 & 35.10 & 2.030 \\
X          & 1.76 & 91.38 & 35.10 & 1.674 \\
Microsoft  & 2.99 & 93.34 & 37.78 & 2.917 \\
Netflix    & 2.10 & 92.90 & 37.10 & 2.040 \\
\hline
\end{tabular}%
}
\label{tab:runtime_evaluation}
\end{table}